\documentclass[]{spie}  
\usepackage[]{graphicx}
\usepackage{subfigure}

\title{The Evryscope: the first full-sky gigapixel-scale telescope}

\author{Nicholas M. Law, Octavi Fors, Philip Wulfken, Jeffrey Ratzloff, and Dustin Kavanaugh
\skiplinehalf
Department of Physics and Astronomy, University of North Carolina at Chapel Hill, Chapel Hill, NC 27599-3255, USA\\
}

\authorinfo{Further author information: Send correspondence to N.M.L.; nmlaw@physics.unc.edu}
 
\begin{document} 
  \maketitle 

\begin{abstract}
Current time-domain wide-field sky surveys generally operate with few-degree-sized fields and take many individual images to cover large sky areas each night. We present the design and project status of the Evryscope (``wide-seer''), which takes a different approach: using an array of 7cm telescopes to form a single wide-field-of-view pointed at every part of the accessible sky simultaneously and continuously. The Evryscope is a gigapixel-scale imager with a 9060 sq. deg. field of view and has an etendue three times larger than the Pan-STARRS sky survey. The system will search for transiting exoplanets around bright stars, M-dwarfs and white dwarfs, as well as detecting microlensing events, nearby supernovae, and gamma-ray burst afterglows. We present the current project status, including an update on the Evryscope prototype telescopes we have been operating for the last three years in the Canadian High Arctic.
\end{abstract}


\section{INTRODUCTION}
\label{sec:intro} 
Most current general time-domain sky surveys (e.g. the Palomar Transient Factory\cite{Law2009}, Pan-STARRS\cite{Kaiser2010}, SkyMapper\cite{Keller2007}, CRTS\cite{Djorgovski2011}, ATLAS\cite{Tonry2011}, and many others) image thousands of square degrees each night in few-degree-wide segments and use large apertures to achieve deep imaging in those areas. The resulting survey is necessarily optimized for events such as supernovae that occur on day-or-longer timescales. However, these surveys are not sensitive to the very diverse class of shorter-timescale objects, including transiting exoplanets, young stellar variability, eclipsing binaries, microlensing planet events, gamma ray bursts, young supernovae, and other exotic transients.

To reach these rare short-timescale events we are building an instrument that takes a different approach: a large array of telescopes covers the entire visible sky in each and every exposure, repeatedly imaging the entire visible sky and co-adding to achieve depth. This technique has been prohibitively expensive up to now because of the extremely large number of pixels required to cover the sky with reasonable sampling, to say nothing of the logistics of building and maintaining the very large numbers of telescopes. However, consumer digital imaging now provides relatively low-cost cameras and large-aperture lenses. Transiting exoplanet surveys with 100-1000 square degree field sizes have been performed with camera lenses (such as SuperWASP\cite{Pollacco2006}, HAT\cite{Bakos2004}, KELT \cite{Pepper2007, Pepper2012} and XO\cite{McCullough2005}), while supernova surveys such as ASAS-SN\cite{Shappee2014} reach greater depths at lower cadences. To reach wider fields of view several groups are experimenting with multiplexing consumer lenses and low-cost cameras, including our group's high-latitude Compound Arctic Telescope Survey (CATS)\cite{Law2012spie,Law2013} concept, and the Pi of the Sky\cite{Zarnecki2011} and Fly's Eye\cite{Csepany2013} instruments. In this paper we present our general approach to covering very wide sky areas with consumer lenses: a robust single-moving-part system that covers up to $\sim$10,000 square degrees with good pixel sampling in a single instrument on a single mount.

\begin{figure}
   \begin{center}
   \begin{tabular}{c}
   \includegraphics[width=0.7\textwidth]{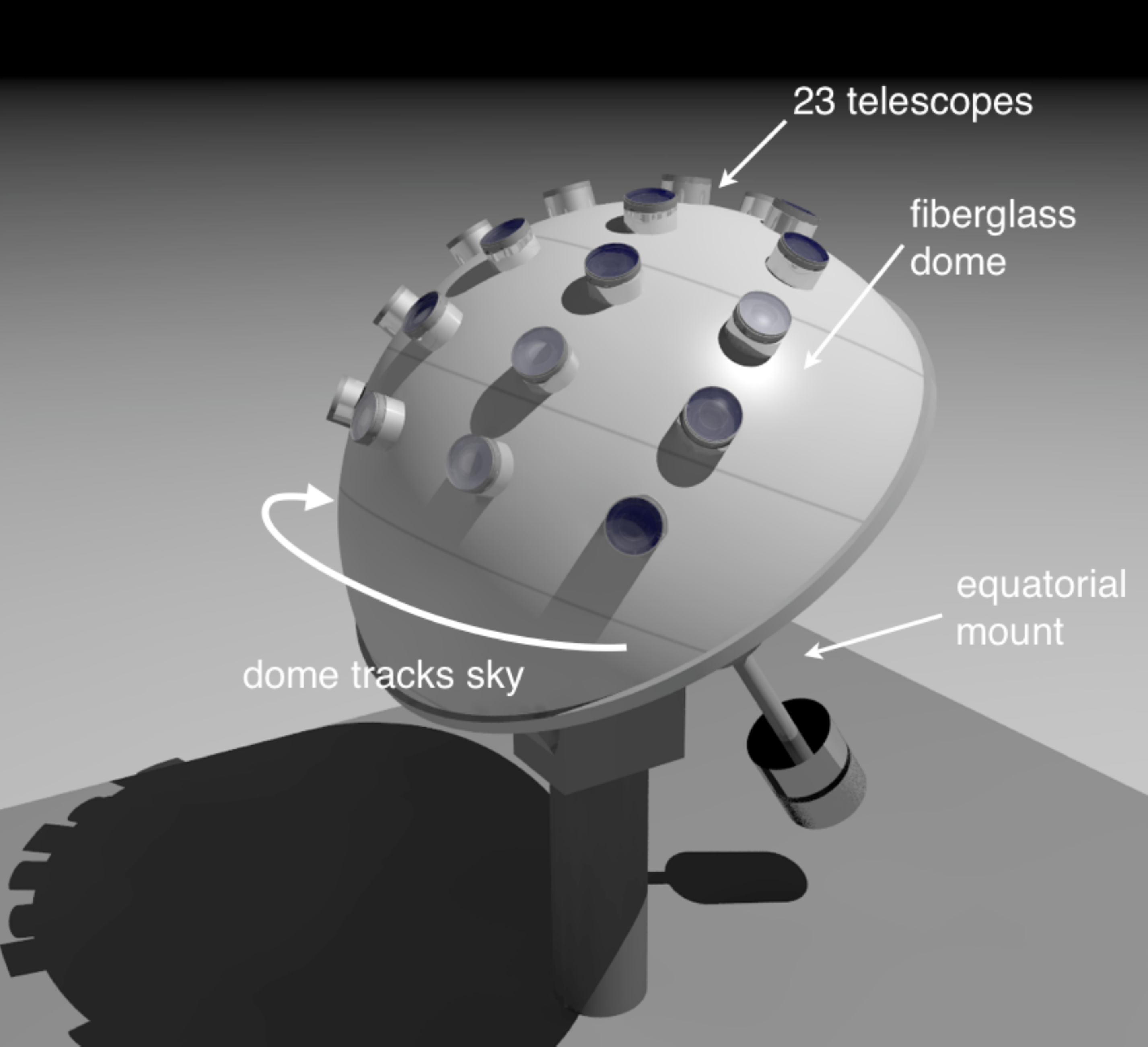}
   \end{tabular}
   \end{center}
   \caption 
   { \label{fig:evryscope} The Evryscope instrument on a German equatorial mount. This example 1.2m-wide dome contains 23 separate 7cm telescopes, delivering a 9060-square-degree instantaneous field of view. The concept easily scales to larger apertures and improved sky sampling.  }
   \end{figure}

Our telescope design (Figure \ref{fig:evryscope}) is essentially a low-cost 0.7 gigapixel robotic telescope which images 9060 square degrees in each exposure. The system forms a 7cm telescope pointed at $\sim$1/4 of the entire sky simultaneously. This capability will allow us to explore the sky in a new way: snapshot images of every visible part of the sky in one go, repeated on minute timescales and co-added over hours and days. Contrasting the traditional telescope (Greek; far-seeing) to this instrument's emphasis on overwhelmingly wide fields, we have coined a name for this new type of telescope: the \textit{Evryscope}, from the Greek for wide-seeing.

We present the Evryscope conceptual design (Section \ref{sec:concept}) and introduce our planned exoplanet, galactic and extragalactic surveys. In Section \ref{sec:prototype} we describe the prototype system we are designing. We detail the project status in Section \ref{sec:status}, along with results from the prototype telescopes we have been operating in the High Canadian Arctic since early 2012.

\section{THE EVRYSCOPE CONCEPT}
\label{sec:concept}
The Evryscope concept mounts an array of individual telescopes into a single hemispherical enclosure (the ``mushroom''; Figure \ref{fig:evryscope}). The array of cameras defines an overlapping grid in the sky that can provide continuous coverage of $>$10,000 square degrees. The camera array is mounted onto an equatorial mount which rotates the mushroom to track the sky with every camera simultaneously for 2-3 hours, before ``ratcheting'' back and starting tracking again on the next sky area. 

Our detailed Evryscope prototype design (Section \ref{sec:prototype}) mounts 23 low-cost telescopes each of which has few-hundred-square-degree fields of view, 29 megapixels, and a 7cm aperture. The Evryscope will take continuous two-minute exposures and co-add to achieve depth. The instrument will allow the detection and monitoring of objects and events as faint as V=16.5 in few-minute exposures and as faint as V=19 after co-adding (the detailed specifications are presented in Section \ref{sec:prototype}).

\subsection{Evryscope Science}

The Evryscope has the potential to open a new parameter space for optical astronomy by trading instantaneous depth and sky sampling for simultaneous coverage of the entire accessible sky. This large dataset will enable extremely-wide-field transiting exoplanet searches, realtime searches for transient and variable phenomena, and it offers the capability to effectively pre-image unexpected events detected by other surveys. Our current designs offer limiting magnitudes that will allow both galactic and extragalactic events to be recorded, and the extremely wide field of view opens the possibility of monitoring large samples of rare objects that up to now have required individual targeting. 

In the following paragraphs we summarize our plans for our prototype Evryscope system. However, with a continuous movie of 1/4 of the entire sky every two minutes, the projects described below only scratch the surface of the system's capabilities.

\textbf{Exoplanet surveys:} Current exoplanet transit surveys are limited to fields of view of 100-1000 square degrees and so cannot effectively search for transits around large samples of stars that occur rarely in the sky. The Evryscope will have an order-of-magnitude larger field of view than the next-largest current exoplanet surveys, which will enable four transiting planet key projects:

\begin{enumerate}
\item{Searching for exoplanets around nearby, bright stars which can be easily followed-up for precision spectrophotometric and mapping techniques\cite{Agol2010,Winn2011,Majeau2012}.}
\item{A habitable-zone survey for rocky transiting planets around bright, nearby M-dwarfs.}
\item{A white-dwarf survey for transiting planets; the Evryscope will be the first extremely-wide-field survey with the time resolution and sensitivity to be able to cover a sample of relatively bright, nearby white-dwarfs by searching for very deep eclipses\cite{Agol2011}.}
\item{TESS precursor observations. The Evryscope will provide long-term monitoring of TESS\cite{Ricker2014} targets, measuring stellar activity and vetting for variable stars. The system will also increase the TESS giant planet yield by recovering multiple transits from objects seen as single eclipses in the relatively short TESS search period.}
\end{enumerate}

The Evryscope prototype will also search for transit and eclipse timing variations to detect non-transiting bodies in transiting exoplanet and eclipsing binary systems. With few-minute cadences over large section of the sky, the Evryscope can also search for and characterize rare gravitational microlensing events involving nearby stars\cite{Han2008}.

\textbf{Variable stars:} The Evryscope prototype will monitor the brightness of millions of stars across the sky each night, building up a multi-year, two-minute-cadence database of stellar activity for every star brighter than V=16.5 in its 110-degree declination range. This will enable the detection and characterization of unprecedented numbers of young and active stars, long-period eclipsing binaries which can be used to constrain the mass/radius relation, as well as the detection of a wide variety of other types of stellar variability.

\textbf{Pre-explosion imaging of gamma-ray-bursts and nearby supernovae.} The Evryscope's short-cadence limit of V=16.5 will enable it to measure the lightcurves of the brightest rapid transients such as gamma-ray-bursts. Co-adding will push the depth to up to V=19 on hour timescales, sufficient to monitor nearby supernovae as they occur. Because all data is recorded, the Evryscope will uniquely be able to provide \textit{pre}-explosion imaging for minutes, weeks or even months before each event. This unique capability will potentially enable the detection of outbursts or the monitoring of rise times from the very beginning of transient events.

\subsection{Sky tracking and camera placement} 

The individual telescopes are mounted in the hemispherical mushroom so that their pointing is set by the normal of the hemisphere's surface (Figure \ref{fig:evryscope}). The mushroom therefore mimics the hemisphere of the visible sky and the rotation of the sky around the Celestial Poles can be tracked on all cameras simultaneously by correspondingly rotating the mushroom. Seen another way, when the mushroom is mounted on an equatorial mount, each camera effectively acts as an individual telescope on the mount, and the system can thus simultaneously track with dozens of cameras using a single moving part.

\begin{figure}
   \begin{center}
   \begin{tabular}{c}
   \includegraphics[width=1.0\textwidth]{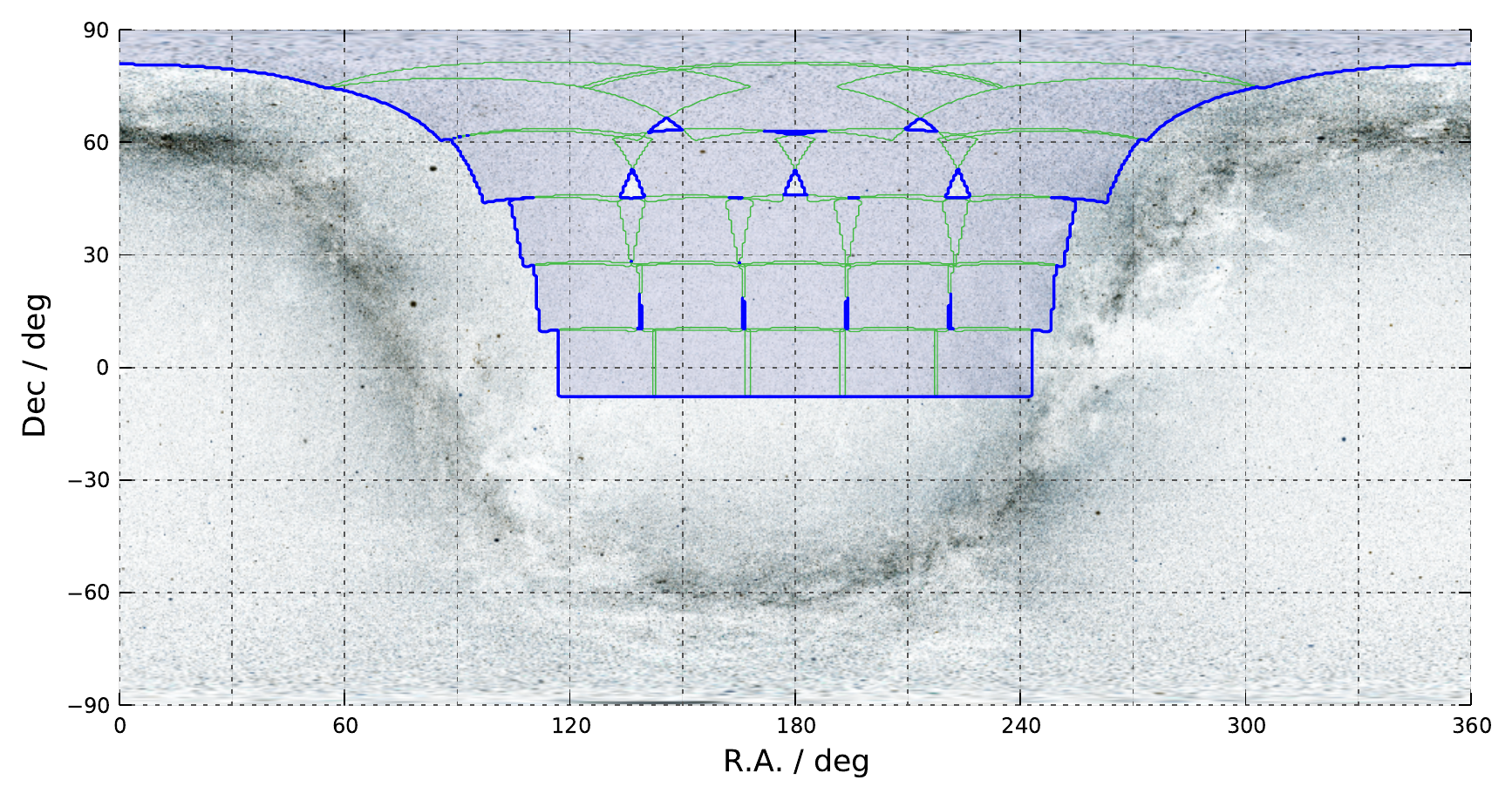}
   \end{tabular}
   \end{center}
   \caption 
   { \label{fig:skygrid} An example Northern-hemisphere Evryscope camera grid arrangement for rectangular camera fields with a $25.4^{\circ}\times18.0^{\circ}$ field of view. This camera placement is optimized for exoplanet transit science by emphasizing Northern fields near the Celestial Pole which can be continuously covered for long periods. The particular R.A. range of the sky covered each night changes throughout the year.}
   \end{figure}

The optimal camera arrangement is a complex function of the camera field of view, the degree of required overlap, and the science requirements. In Figure \ref{fig:skygrid} we show an example camera grid arrangement for a Northern-hemisphere Evryscope with a $\sim$9000-square-degree field of view optimized for an exoplanet transit survey.

\subsection{Operations and Reliability}
The Evryscope is designed to be a completely robotic telescope. Unattended operation and the high chance of failure with dozens of separate camera systems necessitates designing for high reliability and a minimum of moving parts. Our current design uses interline CCDs with an electronic shutter and fixes the filters in place during normal operation. The instrument can thus operate throughout the night with the R.A. drive as the single moving part. The mushroom is sealed to protect against dust and precipitation and our prototype system is designed to operate within a protective AstroHaven clamshell dome.

\section{THE EVRYSCOPE PROTOTYPE}
\label{sec:prototype}
Our prototype Evryscope design consists of a single mushroom (Figure \ref{fig:evryscope}) which contains twenty-three 7cm telescopes, each with a rectangular 28.8MPix interline CCD imaging a 394 sq. deg. FOV using an 85mm f/1.2 or f/1.4 lens.  The dome tracks the sky on a standard German Equatorial mount, with the entire system imaging an instantaneous 9060 sq. deg. FOV. The individual telescopes are fixed into holes in an aluminium-reinforced fibreglass dome and are controlled by six low-power computers fixed into the dome itself. Fans dissipate the 200-300W waste heat. A 5-position filter wheel on each camera allows a selection of filters as well as providing a blocking shutter for dark exposures and daytime light protection for the cameras.

In its 110-degree declination range, we expect that the prototype Evryscope will achieve:

\begin{enumerate}
\item{two-minute-cadence multi-year light curves for every star brighter than V=16.5}
\item{millimagnitude minute-cadence photometry for every star brighter than V=12}
\item{minute-by-minute record of all events in the sky down to V=16.5}
\item{V=19 in one-hour integrations; every part of the sky observed for at least 6.5 hours per night.}
\end{enumerate}

The photometric performance calculations use conservative assumptions: median V-band sky brightness V=21.8; atmosphere + telescope throughput 45\%; 50\% of PSF light landing within a 4-pixel aperture, and a 25\% light-loss due to average vignetting and angular quantum efficiency effects across the field. We have validated our limiting magnitude and photometric performance models using the achieved performance of our already-deployed prototype cameras which use the same lenses and similar CCDs.

For initial operations we have optimized the camera arrangement to maximize the sky area with continuous coverage during nighttime. The Evryscope will track 9060 square degrees for two hours at a time before moving back; the field of view is wide enough that each star is observed for at least 6.5 hours each night (more at higher declinations). This ``ratcheting'' survey setup is designed for both exoplanet transit searches (precision long-term photometry) and co-adding of images for deep imaging (transient searches). The survey maintains pixel-level positioning of stars on two-hour timescales, greatly aiding in both precision photometry and co-adding on those timescales. There will be step-function systematics associated with the two-hour-timescale swap to new cameras, but their removal will be facilitated by the baseline of at least 60 photometric points either side of the swap.

   \begin{figure}
   \begin{center}
   \begin{tabular}{c}
   \includegraphics[width=1.0\textwidth]{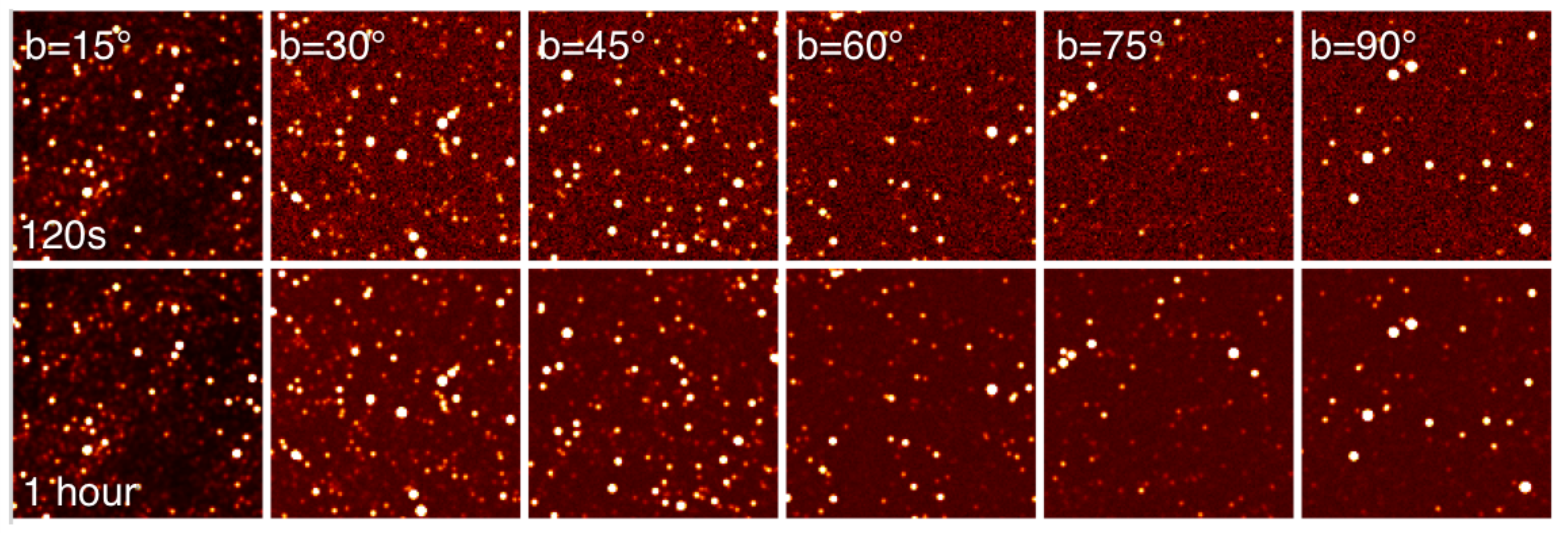}
   \end{tabular}
   \end{center}
   \caption 
   { \label{fig:sampling}We evaluated the crowding levels of Evryscope images by developing a tool to simulate Evryscope images on the basis of the USNO-B1 catalog\cite{Monet2003}, including the effects of crowding, the lens PSFs, the detector sampling, and photon and detector noise. We show above representative 10-arcminute fields in single exposures and in one-hour co-adds, scaled to show the faintest detectable stars. Crowding does not limit photometry for at least 90\% of stars above 15 degrees galactic latitude in 120s exposures (30 degrees in 60m exposures).}
   \end{figure}

The prototype system delivers a pixel sampling of 13"/pixel. Our sky simulations (Figure \ref{fig:sampling}) show that, given the Evryscope limiting magnitude, crowding is only likely to be a limiting factor in the standard two-minute exposures in some sky areas below 15 degrees of galactic latitude (or $<$30 degrees in one-hour coadds).

\subsection{Data handling}
The Evryscope will produce a 660MPix image every 2 minutes, or 5MB/sec of data on average. The data will be stored on network storage units at the telescope; each 20TB unit has storage for approximately three months of data assuming 2/3 good weather. We intend to store all data produced by the Evryscope to enable after-the-fact data mining of events in every visible part of the sky. 

We are developing a pipeline that will produce two data products: 1) photometrically and astrometrically calibrated images; and 2) precision photometric light curves for every star in the Evryscope FOV. We will base our data reduction pipeline on the millimagnitude-precision-photometry system we have already developed for the AWCams and the PTF survey\cite{Law2009,Law2012,Law2013}. The first available data products will include cutout images for interesting sky positions and times, along with precision photometry for selected target lists. We intend to release data to interested astronomers on request.

\section{PROJECT STATUS}
\label{sec:status}

We have been operating two prototypes of the individual Evryscope telescopes at a High Canadian Arctic site since early 2012. The full Evryscope is currently under development and we expect to deploy an early prototype with a reduced camera count by early 2015.

\subsection{Arctic prototype cameras: the AWCams}

The AWCams\cite{Law2013} (Arctic Wide-field Cameras) are two small telescopes designed to search for exoplanet transits around bright stars (V=5-10). The telescopes, with large CCDs behind camera lenses, are essentially identical to individual Evryscope cameras.  In early 2012 we (in collaboration with the University of Toronto; see Law et al. 2012, 2013) deployed the telescopes to the PEARL atmospheric science laboratory at 80$^{\circ}$N in the Canadian High Arctic, where continuous darkness in winter gives a greatly improved transit detection efficiency\cite{Law2012spie,Law2013}.  The cameras and the AWCam project are described in detail elsewhere\cite{Law2012spie, Law2013}.

\begin{figure}
    \subfigure{\resizebox{0.48\textwidth}{!}{{\includegraphics{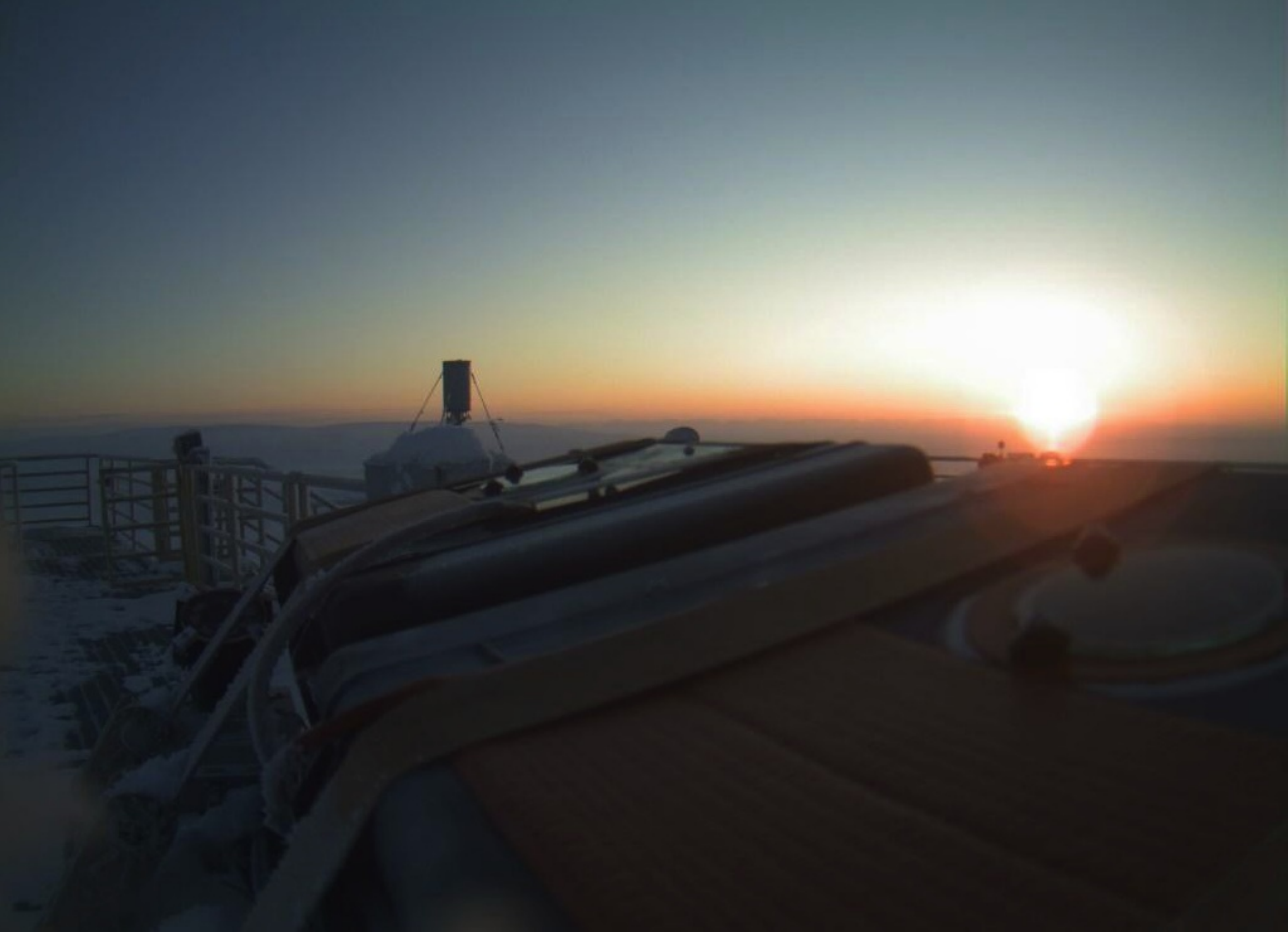}}}}\hspace{0.15in}
    \subfigure{\resizebox{0.45\textwidth}{!}{{\includegraphics{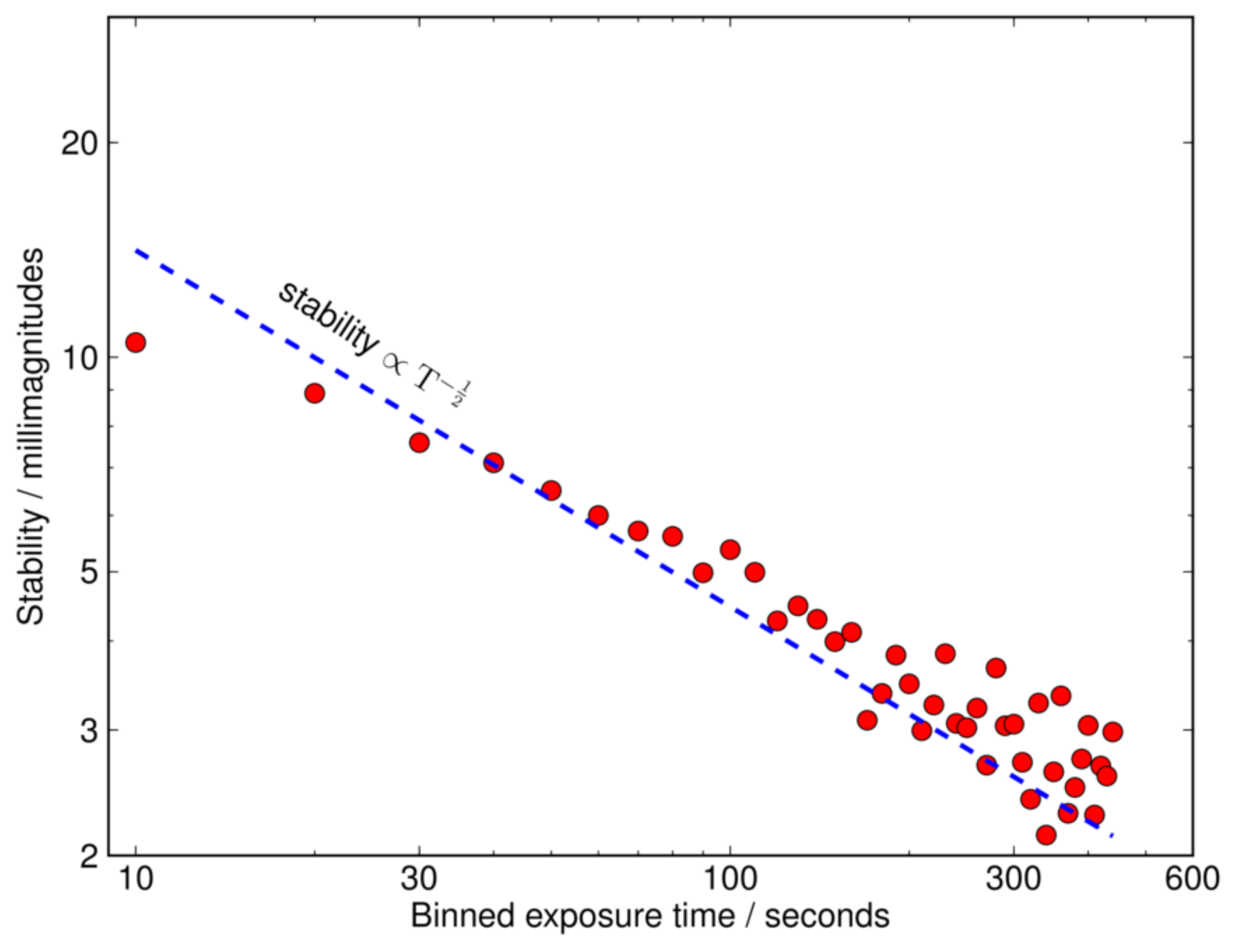}}}}\hspace{0.15in}

   \caption
   { \label{fig:awcams} \textit{Left:} The AWCam Arctic wide-field exoplanet search telescopes and the first sunrise after a full winter of unattended operation.  \textit{Right:} Photometric precision as a function of binning timescale for one of the AWCam telescopes.}
   \end{figure}

Like the Evryscope telescopes, the AWCams (Arctic Wide-field Cameras) are based on Canon f/1.2 lenses (85mm and 50mm focal length) and low-cost front-illuminated CCDs (the 16MPix Kodak KAF16803). The two cameras stare at 500-1000 square degrees around the North Celestial Pole, taking short 10-second exposures to avoid the need to track. The PEARL laboratory is currently unoccupied during the winter months and the AWCams must operate in -40C temperatures and survive Arctic winter storms, which can produce several feet of snow and 100+ mph winds. 

We have now operated the AWCams for three winters, including a test run in February 2012 and full-winter operations in the 2012/13 and 2013/14 winters. The robustness of our hardware and enclosure design has been validated by perfect operation throughout the entire deployment period, including a total of 10 months of completely unattended robotic operation. Throughout the winter the cameras kept themselves (and crucially their windows) clear of snow and ice, took over 40TB of images, and consistently maintained few-millimagnitude photometric precisions (Law et al. 2014, these proceedings). The AWCams photometric performance demonstrates that we can routinely achieve precision photometry with the Evryscope telescopes (Figure \ref{fig:awcams}). The AWCam pipeline achieves scintillation-limited photometric precision in each exposure, and binning brings the photometric precision to the 3-millimagnitude level in 10 minutes.

\subsection{Full Evryscope Deployment}
The full Evryscope design is currently under development at UNC Chapel Hill, including camera unit testing, mechanical design, and the development of the Evryscope software pipeline. The Evryscope's modular construction allows us to scale the number of telescopes as resources become available, and so we can deploy cut-down versions of the system as early prototypes.

\begin{figure}
   \begin{center}
   \begin{tabular}{c}
   \includegraphics[width=1.0\textwidth]{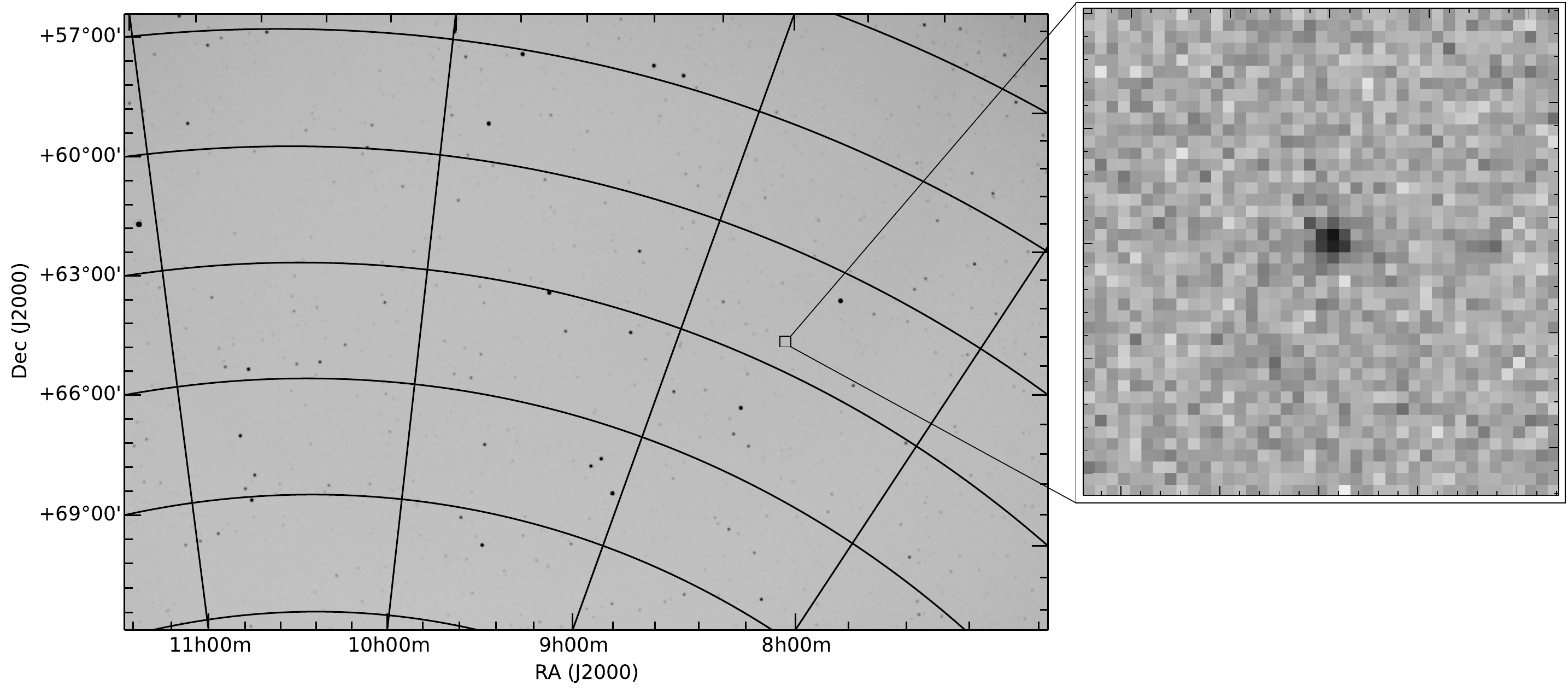}
   \end{tabular}
   \end{center}
   \caption 
   { \label{fig:evryscope_field} The field of view of one of the prototype Evryscope cameras. We have blurred the main image to show stars in this very undersampled figure; an example image of a faint star is shown in an inset to illustrate the image quality of the Rokinon 85mm F/1.4 lens.}
   \end{figure}

We are currently testing individual interline-CCD camera units with encouraging results -- the interline chips are capable of achieving similar photometric precisions to our Arctic cameras, with the advantage of not requiring a mechanical shutter. We are also evaluating the image quality of lower-cost manual focus lenses, compared to the Canon lenses we used in the Arctic. We have found that the Rokinon 85mm F/1.4 lens offers similar image quality (measured in terms of point spread function width and ghosting) to the Canon 85mm F/1.2 lens, at approximately an order of magnitude lower cost (Figure \ref{fig:evryscope_field}). We are currently designing automated methods of adjusting the focus on these manual lenses.

We expect to deploy the prototype system with a reduced camera count in early 2015, most likely to the CTIO observatory in Chile. The prototype will allow us to 1) validate the mushroom design concept; 2) test the reliability and operations of the system; and 3) start our exoplanet-search and transient monitoring programs.

\section{SUMMARY}

Evryscope systems offer a new capability: a continuously recorded image of a large fraction of the visible sky, with the ability to follow individual events and objects on minute-by-minute timescales in archival data.  Our prototype Evryscope system contains twenty-three 7cm telescopes, each with a rectangular 28.8MPix interline CCD imaging a 394 sq. deg. FOV using an 85mm f/1.2 or f/1.4 lens.  The dome tracks the sky on a standard German Equatorial mount, imaging an instantaneous 9060 sq. deg. field of view and forming a low-cost 0.7 gigapixel robotic telescope that images 9060 square degrees in each exposure. 

We have been operating prototypes of the Evryscope's individual telescopes for the last several years in the High Canadian Arctic, taking advantage of the continuous wintertime darkness to search for long-period exoplanets around bright stars. We expect to deploy a prototype of the Evryscope system in early 2015, with a full system following thereafter.

\bibliography{refs3}
\bibliographystyle{spiebib}   

\end{document}